\documentclass[10pt,preprintnumbers,showpacs]{revtex4}
\usepackage{graphicx,epsfig,axodraw}

\def\xslash#1{{\rlap{$#1$}/}}
\def\half{\frac{1}{2}}
\def\beq{\begin{equation}}
\def\eeq{\end{equation}}
\def\beqa{\begin{eqnarray}}
\def\eeqa{\end{eqnarray}}
\def\iar{\begin{array}{l}}
\def\ear{\end{array}}

\begin{document}

\title{Imaginary part of Feynman amplitude, cutting rules and optical theorem}
\author{Yong Zhou}
\affiliation{Institute of High Energy Physics, Academia Sinica, P.O. Box 918(4), Beijing 100049, China, Email: zhouy@ihep.ac.cn}

\begin{abstract}
We discuss the algorithm of the cutting rules of calculating the imaginary part of physical amplitude and the optical theorem. We ameliorate the conventional cutting rules to make it suitable for actual calculation and give the right imaginary part of physical amplitude. The calculations of the imaginary parts of several Feynman diagrams show that the ameliorated cutting rules agrees with the conventional integral algorithm very well. The investigation helps us to find that the optical theorem has severe contradictions and problem, thus isn't right. Besides, the calculation of a physical amplitude show that the ameliorated cutting rules keeps the physical result gauge-parameter independent.
\end{abstract}

\maketitle

\section{Introduction}

It's well known that the discontinuity of a Feynman amplitude comes from the singularity of the Feynman propagator \cite{c0,c1}. In an 1-Particle-Irreducible (1PI) Feynman diagram the discontinuity will contribute imaginary part to the Feynman amplitude. One of the way to calculate the discontinuity is the causal perturbative theory \cite{c1}, another is called as cutting rules proposed by Cutkosky \cite{c0}. Here we mainly discuss the cutting rules. We note that although the cutting rules is well-defined, it hasn't given an integrative algorithm to practically calculate the discontinuity. So we want to investigate it further.

On the other hand, the optical theorem has given a strong constraint on the imaginary part of physical amplitude. Whether they agree with each other hasn't been clearly investigated. So we will discuss their relationship and investigate the origin of the optical theorem. The arrangement of this paper is: firstly we discuss how to ameliorate the cutting rules and investigate its physical meaning; then we calculate the imaginary parts of some Feynman diagrams to see if the ameliorated cutting rules coincides with the conventional integral algorithm; in section IV we investigate the breaking of the optical theorem; in Sect.V we calculate a physical amplitude to see if the physical result keeps gauge independent under the ameliorated cutting rules; lastly we give the conclusion.

\section{Ameliorate the conventional cutting rules}

We usually encounter branch cut in Feynman amplitude calculation. Such branch cut will contribute imaginary part to Feynman amplitude in an 1PI Feynman diagram. Consider the Feynman propagator
\beq
  D_F(x-y)\,=\,\int\frac{d^4 p}{(2\pi)^4}\frac{i}{p^2-m^2+i\varepsilon}
  e^{-i p\cdot(x-y)}\,,
\eeq
there exist two singularities $p_0=\pm(({\bf p}^2+m^2)^{1/2}-i\varepsilon)$ in the denominator. They separately represent the processes of the on-shell particle propagating from position $y$ to position $x$ and from position $x$ to position $y$. The cutting rules is just the algorithm to calculate the contribution of the processes to Feynman amplitude. But it isn't suitable for actual calculation since it assumes that only one of the two singularities of Eq.(1) has contribution to Feynman amplitude, however, it doesn't tell which singularity is the case \cite{c0}. In the follows we investigate this problem from the imaginary part of the unstable particle's self energy. We firstly calculate the imaginary part of the unstable particle's self energy when keeping all of the contributions of the singularities of the Feynman propagators to Feynman amplitude. At one-loop level the result is
\beqa
  Im(-i)\int\frac{d^4 k}{(2\pi)^4}\frac{1}{k^2-m_1^2+i\varepsilon}
  \frac{1}{(k-p)^2-m_2^2+i\varepsilon}&\rightarrow&
  \frac{(m^4+m_1^4+m_2^4-2 m^2 m_1^2-2 m^2 m_2^2-2 m_1^2 m_2^2)^{1/2}}{16\pi m^2}
  \nonumber \\
  &\times&\bigl{(} \theta[m_1-m-m_2]+\theta[m-m_1-m_2]+\theta[m_2-m-m_1] \bigr{)}\,,
\eeqa
where the external-line momentum $p$ is on shell $p_0=({\bf p}^2+m^2)^{1/2}$ and $\theta$ is the Heaviside function. According to the Breit-Wigner formula the imaginary part of the unstable particle's one-loop self energy is proportional to its decay width: $Im {\cal M}(p\rightarrow p)=m\,\Gamma$ \cite{c2}. So one can easily see that the result of Eq.(2) is wrong since only the second term of the right-hand side of Eq.(2) satisfies the Breit-Wigner formula. We can draw a sketch to illustrate the origin of the three terms of the right-hand side of Eq.(2),
\vspace{2mm}
\begin{center} \begin{picture}(278,25)
  \SetScale{1.1} \SetWidth{0.45}
  \ArrowLine(0,0)(25,0)
  \ArrowArcn(38,0)(13,180,0)
  \ArrowArcn(38,0)(13,0,180)
  \ArrowLine(50,0)(75,0)
  \Text(10,8)[]{$m$}
  \Text(42,22)[]{$m_1$}
  \Text(42,-23)[]{$m_2$}
  \ArrowLine(85,0)(110,0)
  \ArrowArc(123,0)(13,180,0)
  \ArrowArcn(123,0)(13,180,0)
  \ArrowLine(135,0)(160,0)
  \Text(104,8)[]{$m$}
  \Text(134,22)[]{$m_1$}
  \Text(134,-23)[]{$m_2$}
  \ArrowLine(170,0)(195,0)
  \ArrowArc(208,0)(13,0,180)
  \ArrowArc(208,0)(13,180,0)
  \ArrowLine(220,0)(245,0)
  \Text(198,8)[]{$m$}
  \Text(230,22)[]{$m_1$}
  \Text(230,-23)[]{$m_2$}
\end{picture} \end{center} \vspace{11mm}
where the arrow denotes the propagator with it is cut and the
momentum $q$ along it satisfies the {\em positive} on-shell
condition $q_0=({\bf q}^2+M^2)^{1/2}$ ($M$ is the mass of the
propagator). We call such momentum propagating direction denoted
by the arrow as {\em positive-on-shell} momentum propagating
direction. The three cut self-energy diagrams separately in turn
represent the origins of the three terms of the right-hand side of
Eq.(2). Obviously only the second cut is acceptable by the
Breit-Wigner formula, the others must be eliminated. Comparing the
three cuts we find that the only difference between them is that
the two {\em positive-on-shell} momentum propagating directions of
the cut propagators of the second cut are inverse in the momentum
loop, but that of the others are equi-directional in the momentum
loop. We can use this point to constrain which cut is acceptable,
i.e. the {\em positive-on-shell} momentum propagating directions
of the cut propagators must be inverse in every cut momentum loop.
In an 1PI Feynman diagram such constraint makes all of the {\em
positive-on-shell} momentum propagating directions of the cut
propagators are along a same direction, i.e. the momentum-energy
propagating direction of the external-line particles (see Fig.1-5
for a rough knowledge). We note that such picture can be used to
describe the on-shell virtual particles' propagating processes
happening in the quantum field vacuum, if such processes exist.
Obviously in this picture the on-shell virtual particles satisfy
the energy conservation law. Contrarily, if the above constraint
doesn't exist, these on-shell virtual particles will not keep the
energy conserved because they will provide with energy
reciprocally in the cut momentum loops thus the total energy of
them can be greater than that of the incoming particles.

Obviously there are at most two cuts in each momentum loop based on this constraint. On the other hand, according to the knowledge that cutting once in a momentum loop is equivalent to performing the conventional loop momentum integral, each cut momentum loop can only be cut twice. For the uncut propagator, since its singularity has no contribution to Feynman amplitude, it should be replaced by its Cauchy principle value in the loop momentum integral. Summing up all the discussions we obtain the following ameliorated cutting rules:
\begin{enumerate}
\item Cut through the Feynman diagram in all possible ways such that the cut propagators can simultaneously be put on mass shell and the cut propagators' {\em positive-on-shell} momentum propagating directions are reverse in every cut momentum loop; keep only two cuts in each cut momentum loop.
\item For each cut propagator with definite {\em positive-on-shell} momentum propagating direction, replace $1/(p^2-m^2+i\varepsilon)\rightarrow -2\pi\,i\,\theta[p_0]\,\delta(p^2-m^2)$ (where the momentum $p$ is along the {\em positive-on-shell} momentum propagating direction), for each uncut propagator, apply Cauchy principle value to it, then perform the loop integrals.
\item Sum the contributions of all possible cuts.
\end{enumerate}

One can easily find that the above ameliorated cutting rules is
different from the conventional one in some aspects. Contrarily to
the conventional cutting rules, the ameliorated one constrains
that only two cuts can exist in one momentum loop and implies that
both of the two singularities of Eq.(1) can have contributions to
Feynman amplitude. Besides, it definitely points out that the
Cauchy principle value should be applied to the uncut propagator
in the loop momentum integrals. We note that to the best of our
knowledge there exist no similar discussions before. But then
there exist some calculations of the imaginary parts of the Green
functions by the causal perturbative theory which coincide with
our results \cite{c3} (see section III).

The above ameliorated cutting rules is very abstract. So we give some illustrations in Fig.1-5. The arrows in Fig.1-5 represent the {\em positive-on-shell} momentum propagating directions of the cut propagators. In order to explain the cutting conditions we give two examples in Fig.2.
\vspace{4mm}
\begin{center} \begin{picture}(218,25)
  \SetScale{1.1} \SetWidth{0.45}
  \ArrowLine(0,0)(25,0)
  \ArrowLine(25,0)(46,13)
  \ArrowLine(46,13)(60,20)
  \ArrowLine(25,0)(46,-13)
  \ArrowLine(46,-13)(60,-20)
  \Line(46,13)(46,-13)
  \ArrowLine(70,0)(95,0)
  \ArrowLine(95,0)(116,13)
  \ArrowLine(116,13)(130,20)
  \Line(95,0)(116,-13)
  \ArrowLine(116,-13)(130,-20)
  \ArrowLine(116,-13)(116,13)
  \ArrowLine(140,0)(165,0)
  \Line(165,0)(186,13)
  \ArrowLine(186,13)(200,20)
  \ArrowLine(165,0)(186,-13)
  \ArrowLine(186,-13)(200,-20)
  \ArrowLine(186,13)(186,-13)
\end{picture} \vspace{10mm} \\
{\small FIG. 1: Cuts of the one-loop three-point irreducible Feynman diagram.}
\end{center}
\vspace{2mm}
\begin{center} \begin{picture}(335,90)
  \ArrowLine(0,0)(25,0)
  \Line(25,0)(50,0)
  \ArrowLine(50,0)(75,0)
  \ArrowLine(25,25)(25,0)
  \Line(50,0)(50,25)
  \ArrowLine(0,25)(25,25)
  \ArrowLine(25,25)(50,25)
  \ArrowLine(50,25)(75,25)
  \ArrowLine(85,0)(110,0)
  \ArrowLine(110,0)(135,0)
  \ArrowLine(135,0)(160,0)
  \ArrowLine(110,0)(110,25)
  \Line(135,0)(135,25)
  \ArrowLine(85,25)(110,25)
  \Line(110,25)(135,25)
  \ArrowLine(135,25)(160,25)
  \ArrowLine(170,0)(195,0)
  \Line(195,0)(220,0)
  \ArrowLine(220,0)(245,0)
  \Line(195,25)(195,0)
  \ArrowLine(220,0)(220,25)
  \ArrowLine(170,25)(195,25)
  \ArrowLine(195,25)(220,25)
  \ArrowLine(220,25)(245,25)
  \ArrowLine(255,0)(280,0)
  \ArrowLine(280,0)(305,0)
  \ArrowLine(305,0)(330,0)
  \Line(280,25)(280,0)
  \ArrowLine(305,25)(305,0)
  \ArrowLine(255,25)(280,25)
  \Line(280,25)(305,25)
  \ArrowLine(305,25)(330,25)
  \ArrowLine(25,55)(50,55)
  \ArrowLine(50,55)(75,55)
  \ArrowLine(75,55)(100,55)
  \Line(50,80)(50,55)
  \Line(75,55)(75,80)
  \ArrowLine(25,80)(50,80)
  \ArrowLine(50,80)(75,80)
  \ArrowLine(75,80)(100,80)
  \ArrowLine(125,55)(150,55)
  \Line(150,55)(175,55)
  \ArrowLine(175,55)(200,55)
  \ArrowLine(150,80)(150,55)
  \ArrowLine(175,80)(175,55)
  \ArrowLine(125,80)(150,80)
  \Line(150,80)(175,80)
  \ArrowLine(175,80)(200,80)
  \Text(133,88)[]{$p_1$}
  \Text(192,88)[]{$p_2$}
  \Text(162,45)[]{$(p_{10}>p_{20})$}
  \ArrowLine(225,55)(250,55)
  \Line(250,55)(275,55)
  \ArrowLine(275,55)(300,55)
  \ArrowLine(250,55)(250,80)
  \ArrowLine(275,55)(275,80)
  \ArrowLine(225,80)(250,80)
  \Line(250,80)(275,80)
  \ArrowLine(275,80)(300,80)
  \Text(233,88)[]{$p_1$}
  \Text(292,88)[]{$p_2$}
  \Text(262,45)[]{$(p_{10}<p_{20})$}
\end{picture} \vspace{6mm} \\
{\small FIG. 2: Cuts of the one-loop four-point irreducible Feynman diagram.}
\end{center}
\vspace{1mm}
\begin{center} \begin{picture}(396,20)
  \SetScale{1.1} \SetWidth{0.45}
  \ArrowLine(0,0)(25,0)
  \ArrowArcn(40,0)(15,180,90)
  \ArrowArc(40,0)(15,180,270)
  \Line(40,15)(40,-15)
  \CArc(40,0)(15,0,90)
  \CArc(40,0)(15,270,360)
  \ArrowLine(55,0)(80,0)
  \SetOffset(-10,0)
  \ArrowLine(95,0)(120,0)
  \ArrowArcn(135,0)(15,90,0)
  \ArrowArc(135,0)(15,270,360)
  \Line(135,15)(135,-15)
  \CArc(135,0)(15,90,180)
  \CArc(135,0)(15,180,270)
  \ArrowLine(150,0)(175,0)
  \SetOffset(-20,0)
  \ArrowLine(190,10)(215,10)
  \Line(215,10)(235,10)
  \ArrowArc(235,10)(20,180,270)
  \ArrowLine(235,10)(235,-10)
  \ArrowArcn(235,-10)(20,90,0)
  \Line(235,-10)(255,-10)
  \ArrowLine(255,-10)(280,-10)
  \SetOffset(-30,0)
  \ArrowLine(295,-10)(320,-10)
  \Line(320,-10)(340,-10)
  \ArrowArcn(340,-10)(20,180,90)
  \ArrowLine(340,-10)(340,10)
  \ArrowArc(340,10)(20,270,360)
  \Line(340,10)(360,10)
  \ArrowLine(360,10)(385,10)
\end{picture} \vspace{10mm} \\
{\small FIG. 3: Cuts of the two-loop two-point irreducible Feynman diagram.}
\end{center}
\vspace{2mm}
\begin{center} \begin{picture}(376,70)
  \SetScale{0.80} \SetWidth{0.625}
  \ArrowLine(0,15)(25,15)
  \ArrowLine(25,15)(55,15)
  \Line(55,15)(80,15)
  \ArrowLine(80,15)(105,15)
  \CArc(55,15)(30,180,270)
  \ArrowLine(55,-15)(55,15)
  \ArrowLine(55,-15)(80,-15)
  \Line(80,-15)(80,15)
  \ArrowLine(80,-15)(105,-15)
  \ArrowLine(120,-15)(145,-15)
  \ArrowLine(145,-15)(175,-15)
  \Line(175,-15)(200,-15)
  \ArrowLine(200,-15)(225,-15)
  \CArc(175,-15)(30,90,180)
  \ArrowLine(175,15)(175,-15)
  \ArrowLine(175,15)(200,15)
  \Line(200,-15)(200,15)
  \ArrowLine(200,15)(225,15)
  \ArrowLine(240,-15)(265,-15)
  \Line(265,-15)(320,-15)
  \ArrowLine(320,-15)(345,-15)
  \ArrowArcn(295,-15)(30,180,90)
  \ArrowLine(295,-15)(295,15)
  \ArrowLine(320,-15)(320,15)
  \Line(295,15)(320,15)
  \ArrowLine(320,15)(345,15)
  \ArrowLine(360,15)(385,15)
  \Line(385,15)(440,15)
  \ArrowLine(440,15)(465,15)
  \ArrowArc(415,15)(30,180,270)
  \ArrowLine(415,15)(415,-15)
  \ArrowLine(440,15)(440,-15)
  \Line(415,-15)(440,-15)
  \ArrowLine(440,-15)(465,-15)
  \ArrowLine(30,65)(55,65)
  \ArrowLine(55,65)(76,78)
  \Line(76,78)(90,85)
  \ArrowLine(90,85)(104,92)
  \ArrowLine(55,65)(76,52)
  \Line(76,52)(90,45)
  \ArrowLine(90,45)(104,38)
  \Line(76,78)(76,52)
  \Line(90,85)(90,45)
  \ArrowLine(140,65)(165,65)
  \Line(165,65)(186,78)
  \ArrowLine(186,78)(200,85)
  \ArrowLine(200,85)(214,92)
  \Line(165,65)(186,52)
  \ArrowLine(186,52)(200,45)
  \ArrowLine(200,45)(214,38)
  \Line(186,78)(186,52)
  \Line(200,85)(200,45)
  \ArrowLine(250,65)(275,65)
  \Line(275,65)(296,78)
  \ArrowLine(296,78)(310,85)
  \ArrowLine(310,85)(324,92)
  \Line(275,65)(296,52)
  \Line(296,52)(310,45)
  \ArrowLine(310,45)(324,38)
  \Line(296,78)(296,52)
  \ArrowLine(310,45)(310,85)
  \ArrowLine(360,65)(385,65)
  \Line(385,65)(406,78)
  \Line(406,78)(420,85)
  \ArrowLine(420,85)(434,92)
  \Line(385,65)(406,52)
  \ArrowLine(406,52)(420,45)
  \ArrowLine(420,45)(434,38)
  \Line(406,78)(406,52)
  \ArrowLine(420,85)(420,45)
  \SetScale{1.}
\end{picture} \vspace{10mm} \\
{\small FIG. 4: Cuts of the two-loop three-point irreducible Feynman diagram.}
\end{center}
\begin{center} \begin{picture}(326,90)
  \SetScale{0.9} \SetWidth{0.55}
  \ArrowLine(0,0)(30,0)
  \ArrowLine(30,0)(50,0)
  \Line(50,0)(70,0)
  \ArrowLine(70,0)(100,0)
  \ArrowLine(50,20)(50,0)
  \Line(50,0)(50,-20)
  \CArc(50,0)(20,90,180)
  \ArrowArcn(50,0)(20,90,0)
  \ArrowArc(50,0)(20,180,270)
  \CArc(50,0)(20,270,360)
  \ArrowLine(130,0)(160,0)
  \Line(160,0)(180,0)
  \ArrowLine(180,0)(200,0)
  \ArrowLine(200,0)(230,0)
  \Line(180,20)(180,0)
  \ArrowLine(180,0)(180,-20)
  \CArc(180,0)(20,90,180)
  \ArrowArcn(180,0)(20,90,0)
  \ArrowArc(180,0)(20,180,270)
  \CArc(180,0)(20,270,360)
  \ArrowLine(260,0)(290,0)
  \Line(290,0)(310,0)
  \ArrowLine(310,0)(330,0)
  \ArrowLine(330,0)(360,0)
  \ArrowLine(310,0)(310,20)
  \Line(310,0)(310,-20)
  \ArrowArcn(310,0)(20,180,90)
  \CArc(310,0)(20,0,90)
  \CArc(310,0)(20,180,270)
  \ArrowArc(310,0)(20,270,360)
  \ArrowLine(0,60)(30,60)
  \ArrowLine(30,60)(50,60)
  \Line(50,60)(70,60)
  \ArrowLine(70,60)(100,60)
  \Line(50,80)(50,40)
  \ArrowArcn(50,60)(20,180,90)
  \CArc(50,60)(20,0,90)
  \ArrowArc(50,60)(20,180,270)
  \CArc(50,60)(20,270,360)
  \ArrowLine(130,60)(160,60)
  \Line(160,60)(180,60)
  \ArrowLine(180,60)(200,60)
  \ArrowLine(200,60)(230,60)
  \Line(180,80)(180,40)
  \CArc(180,60)(20,90,180)
  \ArrowArcn(180,60)(20,90,0)
  \CArc(180,60)(20,180,270)
  \ArrowArc(180,60)(20,270,360)
  \ArrowLine(260,60)(290,60)
  \ArrowLine(290,60)(310,60)
  \Line(310,60)(330,60)
  \ArrowLine(330,60)(360,60)
  \Line(310,60)(310,80)
  \ArrowLine(310,40)(310,60)
  \ArrowArcn(310,60)(20,180,90)
  \CArc(310,60)(20,0,90)
  \CArc(310,60)(20,180,270)
  \ArrowArc(310,60)(20,270,360)
\end{picture} \\ \vspace{11mm}
{\small FIG. 5: Cuts of the four-loop two-point irreducible Feynman diagram.}
\end{center}
Since each cut contributes an imaginary factor $i$ to Feynman
amplitude, all of the cut diagrams in Fig.1-5 contribute imaginary
parts to the Feynman amplitudes. This agrees with the conventional
knowledge. On the other hand, one can easily find that many
possible cuts in the conventional cutting rules have been
eliminated by the ameliorated one. For example all of the
propagators in Fig.1 and 2 can be simultaneously cut in the
conventional cutting rules \cite{c0}, but this is forbidden in the
ameliorated one. One can also see that both of the two
singularities of many cut propagators have contributions to the
Feynman amplitudes (e.g. the vertical propagator in Fig.1).
Besides, some of the cuts in Fig.1-5 have been used to calculate
the imaginary parts of the Feynman amplitudes \cite{c3,c4}.

\section{Comparison of cutting rules and conventional integral algorithm}

It's well known that the imaginary part of Feynman amplitude can be directly calculated by the conventional integral algorithm, i.e. by the Feynman parametrization, wick rotation and dimensional regularization \cite{c5}. In this section we will give four examples to see whether the results obtained by the ameliorated cutting rules agree with the ones obtained by the conventional integral algorithm.

Firstly we calculate the imaginary part of the two-loop two-point irreducible Feynman diagram
\vspace{4mm}
\begin{center} \begin{picture}(80,20)
\SetScale{1.2} \SetWidth{0.40}
  \ArrowLine(0,2)(25,2)
  \Line(25,2)(50,2)
  \ArrowLine(50,2)(75,2)
  \CArc(38,2)(13,0,180)
  \CArc(38,2)(13,180,0)
  \Text(14,8)[]{\small $p$}
  \Text(45,24)[]{\small $m_1$}
  \Text(45,8)[]{\small $m_1$}
  \Text(45,-8)[]{\small $m_1$}
\end{picture} \end{center}
\vspace{5mm}
where $p^2=m^2$. Using the conventional integral algorithm we have
\beqa
  &&Im\int\frac{d^D k_1}{(2\pi)^D}
  \frac{d^D k_2}{(2\pi)^D}\frac{1}{k_1^2-m_1^2+i\varepsilon}
  \frac{1}{k_2^2-m_1^2+i\varepsilon}\frac{1}{(k_1+k_2-p)^2-m_1^2+i\varepsilon}
  \nonumber \\
  =\hspace{-3mm}&&\frac{1}{256\pi^4}Im\int_0^1 d\,x\int_0^1 d\,y\,
  (A-2 m^2 y)(\frac{2}{\epsilon}-2\gamma+\ln 16\pi^2+\ln y-\ln x(1-x)-\ln B)\ln B
  \nonumber \\
  =\hspace{-3mm}&&-\frac{1}{256\pi^3}\int_0^1 d\,x\int_0^1 d\,y\,
  (A-2 m^2 y)(\frac{2}{\epsilon}-2\gamma+\ln 16\pi^2+\ln y-\ln x(1-x)-2\ln|B|)
  \theta[-B] \nonumber \\
  =\hspace{-3mm}&&\frac{1}{128\pi^3}\int_{x_1}^{x_2}d\,x\Bigl{(}
  \frac{A(A^2-4 m^2 m_1^2)^{1/2}}{4 m^2}-m_1^2\ln\frac{A+(A^2-4 m^2 m_1^2)^{1/2}}{2 m m_1}
  \Bigr{)}\theta[m-3 m_1]\,,
\eeqa
where $\epsilon=4-D$, $\gamma$ is the Euler constant, and
\beqa
  A\,=\hspace{-3mm}&&m^2+m_1^2-\frac{m_1^2}{x(1-x)}\,, \nonumber \\
  B\,=\hspace{-3mm}&&m^2 y^2-A\,y+m_1^2\,, \nonumber \\
  x_{1,2}\,=\hspace{-3mm}&&\frac{m-m_1\mp(m^2-2 m m_1-3 m_1^2)^{1/2}}{2(m-m_1)}\,.
\eeqa
In Eq.(3) we have used the formula $\ln a=\ln|a|-i\pi\,\theta[-a]$. On the other hand, using the ameliorated cutting rules we also have
\beqa
  &&Im\int\frac{d^4 k_1}{(2\pi)^4}\frac{d^4 k_2}
  {(2\pi)^4}\frac{1}{k_1^2-m_1^2+i\varepsilon}\frac{1}{k_2^2-m_1^2+i\varepsilon}\frac{1}
  {(k_1+k_2-p)^2-m_1^2+i\varepsilon} \nonumber \\
  =\hspace{-3mm}&&\frac{1}{64\pi^5}\int d^4 k_1 d^4 k_2\,\theta[k_{10}]\,\delta(k_1^2-m_1^2)
  \,\theta[k_{20}]\,\delta(k_2^2-m_1^2)\,\theta[p_0-k_{10}-k_{20}]\,\delta((p-k_1-k_2)^2-m_1^2) \nonumber \\
  =\hspace{-3mm}&&\frac{1}{64 m\pi^3}\int_{m_1}^{\frac{m^2-3 m_1^2}{2 m}}d\,x
  \Bigl{[}(m x+m_1^2)^2-\frac{m_1^2(m^2-m_1^2)^2}{m^2+m_1^2-2 m x}\Bigr{]}^{1/2}\theta[m-3m_1]\,.
\eeqa
Through numerical calculations we find that Eq.(5) is equal to Eq.(3). The concrete results are shown in Fig.6.
\setcounter{figure}{5}
\begin{figure}[htbp]
\begin{center}
  \epsfig{file=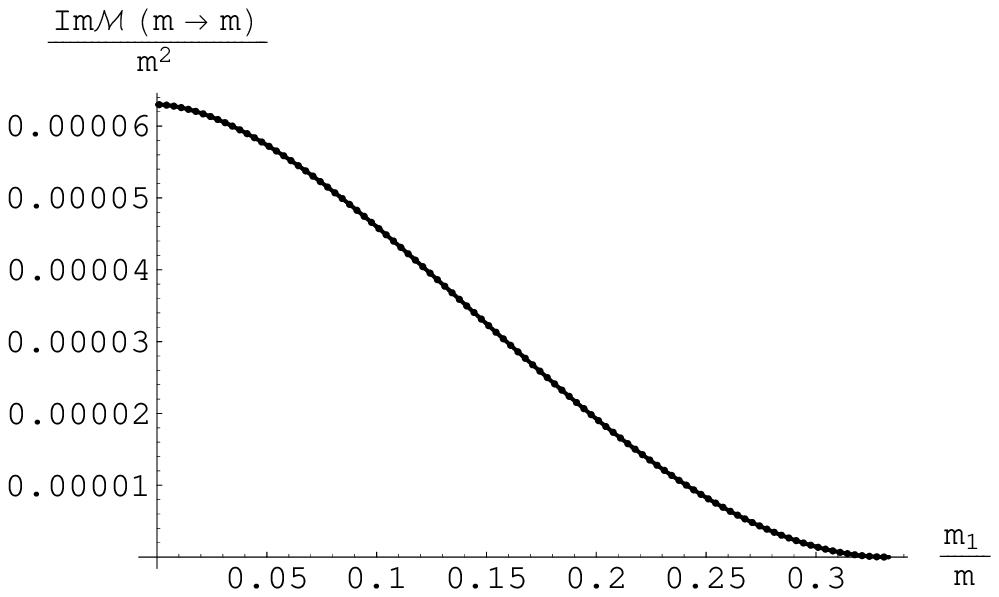,width=7cm} \\
  \caption{Results of Eq.(3) and Eq.(5).}
\end{center}
\end{figure}

Secondly we calculate the imaginary part of the one-loop three-point irreducible Feynman diagram
\vspace{4mm}
\begin{center} \begin{picture}(70,25)
  \SetScale{1.1} \SetWidth{0.45}
  \ArrowLine(0,0)(25,0)
  \Line(25,0)(46,13)
  \Line(25,0)(46,-13)
  \ArrowLine(46,13)(60,20)
  \ArrowLine(46,-13)(60,-20)
  \Line(46,13)(46,-13)
  \Text(10,6)[]{$p$}
  \Text(31,12)[]{$m_1$}
  \Text(31,-12)[]{$m_1$}
  \Text(60,0)[]{$m_1$}
  \Text(74,20)[]{$p_1$}
  \Text(74,-20)[]{$p_2$}
\end{picture} \end{center}
\vspace{8mm}
where $p^2=(p_1+p_2)^2=m_a^2$ and $p_1^2=p_2^2=m_b^2$. Using the conventional integral algorithm we have
\beqa
  &&Im(-i)\int\frac{d^4 k}{(2\pi)^4}\frac{1}{k^2-m_1^2+i\varepsilon}
  \frac{1}{(k-p_1)^2-m_1^2+i\varepsilon}\frac{1}{(k-p_1-p_2)^2-m_1^2+i\varepsilon}
  \nonumber \\
  =\hspace{-3mm}&&-\frac{1}{16\pi^2}\,Im\int_0^1 d\,x\int_0^{1-x} \frac{d\,y}
  {m_a^2\,x^2+m_b^2\,y^2+m_a^2\,x\,y-m_a^2\,x-m_b^2\,y+m_1^2-i\varepsilon} \nonumber \\
  =\hspace{-3mm}&&-\frac{1}{16\pi}\int_0^1 d\,x\int_0^{1-x}d\,y\,
  \delta(m_a^2\,x^2+m_b^2\,y^2+m_a^2\,x\,y-m_a^2\,x-m_b^2\,y+m_1^2) \nonumber \\
  =\hspace{-3mm}&&\left\{ \begin{array}{l}
  \frac{1}{16\pi m_a(m_a^2-4 m_b^2)^{1/2}}\ln \biggl{[} \frac{m_a^2-2 m_b^2-
  \sqrt{(m_a^2-4 m_b^2)(m_a^2-4 m_1^2)}}{m_a^2-2 m_b^2+
  \sqrt{(m_a^2-4 m_b^2)(m_a^2-4 m_1^2)}}\left( \frac{m_a m_b+
  \sqrt{(m_a^2-4 m_b^2)(m_b^2-4 m_1^2)}}{m_a m_b-
  \sqrt{(m_a^2-4 m_b^2)(m_b^2-4 m_1^2)}} \right)^2 \biggr{]}
  \hspace{4mm} \forall\, m_1 \in (0, \frac{m_b}{2} ) \,,\\
  \frac{1}{16\pi m_a\sqrt{m_a^2-4 m_b^2}}\ln\frac{m_a^2-2 m_b^2-
  \sqrt{(m_a^2-4 m_b^2)(m_a^2-4 m_1^2)}}{m_a^2-2 m_b^2+
  \sqrt{(m_a^2-4 m_b^2)(m_a^2-4 m_1^2)}}
  \hspace{56mm} \forall\, m_1\in  (\frac{m_b}{2},\frac{m_a}{2})
  \,, \\ \vspace{1mm} \\   0
  \hspace{127mm} \forall\, m_1\in  (\frac{m_a}{2},\infty)  \,,
  \ear \right.
\eeqa
where the mathematical formula ($P$ denotes the Cauchy principle value)
\beq
  \frac{1}{a\pm i\varepsilon}\,=\,P\frac{1}{a}\mp i\pi\delta(a)
\eeq
has been used in the Feynman parameter integrals. On the other hand, using the ameliorated cutting rules we also have (see Fig.1)
\beqa
  &&Im(-i)\int\frac{d^4 k}{(2\pi)^4}\frac{1}{k^2-m_1^2+i\varepsilon}
  \frac{1}{(k-p_1)^2-m_1^2+i\varepsilon}\frac{1}{(k-p_1-p_2)^2-m_1^2+i\varepsilon}
  \nonumber \\
  =\hspace{-3mm}&&P\int\frac{d^4 k}{8\pi^2}\biggl{[} \frac{\theta[k_0]\delta(k^2-m_1^2)
  \theta[p_{10}+p_{20}-k_0]\delta((p_1+p_2-k)^2-m_1^2)}{(k-p_1)^2-m_1^2+i\varepsilon} \nonumber \\
  +\hspace{-3mm}&&\frac{\theta[k_0]\delta(k^2-m_1^2)\theta[p_{10}-k_0]\delta((p_1-k)^2-m_1^2)}
  {(k-p_1-p_2)^2-m_1^2+i\varepsilon} \nonumber \\
  +\hspace{-3mm}&&\frac{\theta[k_0-p_{10}]\delta((k-p_1)^2-m_1^2)\theta[p_{10}+p_{20}-k_0]
  \delta((p_1+p_2-k)^2-m_1^2)}{k^2-m_1^2+i\varepsilon} \biggr{]} \nonumber \\
  =\hspace{-3mm}&&\frac{\theta[m_a-2 m_1]}{16\pi m_a\sqrt{m_a^2-4 m_b^2}}
  \ln\left|\frac{m_a^2-2m_b^2-\sqrt{(m_a^2-4 m_b^2)(m_a^2-4 m_1^2)}}{m_a^2-2m_b^2+
  \sqrt{(m_a^2-4 m_b^2)(m_a^2-4 m_1^2)}}\right| \nonumber \\
  +\hspace{-3mm}&&\frac{\theta[m_b-2m_1]}{8\pi m_a\sqrt{m_a^2-4 m_b^2}}
  \ln\left|\frac{m_a m_b+\sqrt{(m_a^2-4 m_b^2)(m_b^2-4 m_1^2)}}
  {m_a m_b-\sqrt{(m_a^2-4 m_b^2)(m_b^2-4 m_1^2)}}\right|\,.
\eeqa
Obviously Eq.(8) is equal to Eq.(6). Eq.(8) also coincides with Eq.(3.24) of Ref.\cite{c3}.

Thirdly we calculate the imaginary part of the one-loop four-point irreducible Feynman diagram
\vspace{2mm}
\begin{center} \begin{picture}(100,35)
  \ArrowLine(0,0)(25,0)
  \Line(25,0)(50,0)
  \ArrowLine(50,0)(75,0)
  \Line(25,0)(25,25)
  \Line(50,0)(50,25)
  \ArrowLine(0,25)(25,25)
  \Line(25,25)(50,25)
  \ArrowLine(50,25)(75,25)
  \Text(38,31)[]{$m$}
  \Text(38,-6)[]{$m$}
  \Text(18,13)[]{$m$}
  \Text(57,13)[]{$0$}
  \Text(-6,25)[]{$r$}
  \Text(-6,0)[]{$q$}
  \Text(81,25)[l]{$r=(r_0,p\,{\bf e_z})$}
  \Text(81,0)[l]{$q=(q_0,-p\,{\bf e_z})$}
\end{picture} \end{center} \vspace{3mm}
where $r_0=q_0=2 p$. Using the conventional integral algorithm and Eq.(7) we have
\beqa
  &&Im(-i)\int\frac{d^4 k}{(2\pi)^4}\frac{1}{k^2+i\varepsilon}
  \frac{1}{(k-q)^2-m^2+i\varepsilon}\frac{1}{k^2-m^2+i\varepsilon}
  \frac{1}{(k+r)^2-m^2+i\varepsilon} \nonumber \\
  =\hspace{-3mm}&&\frac{1}{16\pi^2}\,Im\int_0^1 d\,x\int_0^{1-x}d\,y\int_0^{1-x-y}
  \frac{d\,z}{(3 p^2\,x^2+3 p^2\,y^2-10 p^2\,x\,y-3 p^2\,x-3 p^2\,y-m^2\,z+m^2
  -i\varepsilon)^2} \nonumber \\
  =\hspace{-3mm}&&\frac{1}{16\pi^2 m^2}\,Im\int_0^1 d\,x\int_0^{1-x}d\,y\,\left[
  \frac{1}{3 p^2\,x^2+3 p^2\,y^2-10 p^2\,x\,y+(m^2-3 p^2)x+(m^2-3 p^2)y-i\varepsilon}
  \right. \nonumber \\
  -\hspace{-3mm}&&\frac{1}{3 p^2\,x^2+3 p^2\,y^2-10 p^2\,x\,y-3 p^2\,x-3 p^2\,y+m^2
  -i\varepsilon} \biggr{]} \nonumber \\
  =\hspace{-3mm}&&\frac{1}{16\pi m^2}\int_0^1 d\,x\int_0^{1-x}d\,y\,\left[
  \delta(3 p^2\,x^2+3 p^2\,y^2-10 p^2\,x\,y+(m^2-3 p^2)x+(m^2-3 p^2)y)
  \right. \nonumber \\
  -\hspace{-3mm}&&\delta(3 p^2\,x^2+3 p^2\,y^2-10 p^2\,x\,y-3 p^2\,x-3 p^2\,y+m^2) \bigr{]}
  \nonumber \\
  =\hspace{-3mm}&&\frac{\theta[p-\frac{m}{2}]\theta[\frac{m}{\sqrt{3}}-p]}{128\pi m^2 p^2}
  \ln\frac{9p^3-m^2 p+2m^2\sqrt{4p^2-m^2}}{9p^3-m^2 p-2m^2\sqrt{4p^2-m^2}} \nonumber \\
  +\hspace{-3mm}&&\frac{\theta[p-\frac{m}{\sqrt{3}}]\theta[\frac{2m}{\sqrt{3}}-p]}{128\pi m^2 p^2}
  \ln\frac{9p^3-m^2 p+2m^2\sqrt{4p^2-m^2}}{9(9p^3-m^2 p-2m^2\sqrt{4p^2-m^2})} \nonumber \\
  +\hspace{-3mm}&&\frac{\theta[p-\frac{2m}{\sqrt{3}}]}{128\pi m^2 p^2}
 \ln\frac{(9p^3-m^2 p+2m^2\sqrt{4p^2-m^2})(15p^2+4p\sqrt{9p^2-12m^2}-4m^2)}
  {9(9p^3-m^2 p-2m^2\sqrt{4p^2-m^2})(15p^2-4p\sqrt{9p^2-12m^2}-4m^2)} \,.
\eeqa
On the other hand, using the ameliorated cutting rules we also have (see Fig.2)
\beqa
  &&Im(-i)\int\frac{d^4 k}{(2\pi)^4}\frac{1}{k^2+i\varepsilon}
  \frac{1}{(k-q)^2-m^2+i\varepsilon}\frac{1}{k^2-m^2+i\varepsilon}
  \frac{1}{(k+r)^2-m^2+i\varepsilon} \nonumber \\
  =\hspace{-3mm}&&\frac{1}{8\pi^2}P\int d^4 k \Bigl{[} \frac{\theta[k_0]\delta(k^2-m^2)
  \theta[q_0+r_0-k_0]\delta((q+r-k)^2-m^2)}{(k-r)^2((k-r)^2-m^2)} \nonumber \\
  +\hspace{-3mm}&&\frac{\theta[k_0]\delta(k^2-m^2)\theta[r_0-k_0]\delta((r-k)^2-m^2)}
  {k^2((k+q)^2-m^2)}+\frac{\theta[k_0]\delta(k^2-m^2)\theta[q_0-k_0]\delta((q-k)^2-m^2)}
  {k^2((k+r)^2-m^2)} \nonumber \\
  +\hspace{-3mm}&&\frac{\theta[k_0]\delta(k^2)\theta[r_0-k_0]\delta((r-k)^2-m^2)}
  {(k^2-m^2)((k+q)^2-m^2)}+\frac{\theta[k_0]\delta(k^2)\theta[q_0-k_0]\delta((q-k)^2-m^2)}
  {(k^2-m^2)((k+r)^2-m^2)} \Bigr{]} \nonumber \\
  =\hspace{-3mm}&&\frac{\theta[p-m/2]}{128\pi m^2 p^2}\ln\frac{9p^3-m^2 p+2m^2
  \sqrt{4p^2-m^2}}{9p^3-m^2 p-2m^2\sqrt{4p^2-m^2}} \nonumber \\
  +\hspace{-3mm}&&\frac{\theta[p-2m/\sqrt{3}]}{128\pi m^2 p^2}\ln
  \frac{15p^2-4m^2+4p\sqrt{9p^2-12m^2}}{15p^2-4m^2-4p\sqrt{9p^2-12m^2}}
  -\frac{\theta[p-m/\sqrt{3}]}{64\pi m^2 p^2}\ln3\,.
\eeqa
Obviously Eq.(10) is equal to Eq.(9).

Lastly we calculate the imaginary part of a little more complex Feynman diagram
\begin{center} \begin{picture}(380,36)
  \SetScale{0.9} \SetWidth{0.55}
  \ArrowLine(0,0)(30,0)
  \ArrowLine(70,0)(100,0)
  \CArc(50,0)(20,0,65)
  \BCirc(50,20){9}
  \CArc(50,0)(20,115,180)
  \CArc(50,0)(20,180,0)
  \Text(10,8)[]{$p$}
  \Text(26,14)[]{0}
  \Text(46,32)[]{$m$}
  \Text(46,5)[]{$m$}
  \Text(46,-25)[]{$m$}
  \Text(64,14)[]{$m$}
  \Text(100,0)[]{$\sim$}
  \SetOffset(110,0)
  \ArrowLine(0,0)(30,0)
  \ArrowLine(70,0)(100,0)
  \CArc(50,0)(20,0,65)
  \ArrowArcn(50,20)(9,180,0)
  \ArrowArc(50,20)(9,180,0)
  \CArc(50,0)(20,115,180)
  \ArrowArc(50,0)(20,180,0)
  \Text(100,0)[]{$+$}
  \SetOffset(100,0)
  \ArrowLine(130,0)(160,0)
  \ArrowLine(200,0)(230,0)
  \ArrowArcn(180,0)(20,180,115)
  \BCirc(180,20){9}
  \CArc(180,0)(20,0,65)
  \ArrowArc(180,0)(20,180,0)
  \Text(217,0)[]{$+$}
  \SetOffset(90,0)
  \ArrowLine(260,0)(290,0)
  \ArrowLine(330,0)(360,0)
  \CArc(310,0)(20,115,180)
  \BCirc(310,20){9}
  \ArrowArcn(310,0)(20,65,0)
  \ArrowArc(310,0)(20,180,0)
\end{picture} \end{center}
\vspace{9mm}
where $p^2=M^2$ and the cut diagrams have been shown. Using the conventional integral algorithm and Eq.(7) we have
\beqa
  &&Im\,\mu^{2\epsilon}\int\frac{d^D k_1}{(2\pi)^D}\frac{d^D k_2}{(2\pi)^D}
  \frac{1}{k_1^2+i\varepsilon}\frac{1}{k_2^2-m^2+i\varepsilon}
  \frac{1}{(k_2-k_1)^2-m^2+i\varepsilon}\frac{1}{k_1^2-m^2+i\varepsilon}
  \frac{1}{(k_1-p)^2-m^2+i\varepsilon} \nonumber \\
  =\hspace{-3mm}&&\frac{1}{256\pi^4}Im\int_0^1 d\,x\int_0^1 d\,y\int_0^{1-y}
  \frac{d\,z}{M^2 y^2+(m^2-M^2)y+m^2 z-i\varepsilon}\Bigl{[} \Delta+
  \ln\frac{1-y-z}{x-x^2} \nonumber \\
  -\hspace{-3mm}&&\ln(\frac{M^2 y^2}{m^2}-\frac{M^2 y}{m^2}+y+z
  +\frac{1-y-z}{x-x^2}-i\varepsilon)-\ln(\frac{M^2 y^2}{m^2}-\frac{M^2 y}{m^2}+y+z
  -i\varepsilon) \Bigr{]} \nonumber \\
  =\hspace{-3mm}&&\frac{1}{256\pi^3}\int_0^1 d\,x\int_0^1 d\,y\int_0^{1-y}d\,z\,
  \delta(M^2 y^2+(m^2-M^2)y+m^2 z)\Bigl{[} \Delta+\ln\frac{1-y-z}{x-x^2} \nonumber \\
  -\hspace{-3mm}&&\ln|\frac{M^2 y^2}{m^2}-\frac{M^2 y}{m^2}+y+z
  +\frac{1-y-z}{x-x^2}| \Bigr{]} \nonumber \\
  +\hspace{-3mm}&&\frac{1}{256\pi^3}P\int_0^1 d\,x\int_0^1 d\,y\int_0^{1-y}d\,z\,\frac{
  \theta[\frac{M^2 y}{m^2}-\frac{M^2 y^2}{m^2}-y-z-\frac{1-y-z}{x-x^2}]}
  {M^2 y^2+(m^2-M^2)y+m^2 z} \nonumber \\
  +\hspace{-3mm}&&\frac{1}{512\pi^3 m^2}Im\int_0^1 d\,x\int_0^1 d\,y \Bigl{[}
  \ln^2(\frac{M^2 y^2}{m^2}-\frac{M^2 y}{m^2}+y-i\varepsilon)-
  \ln^2(\frac{M^2 y^2}{m^2}-\frac{M^2 y}{m^2}+1-i\varepsilon) \Bigr{]} \nonumber \\
  =\hspace{-3mm}&&\frac{\theta[M-m]\,\theta[2m-M](M^2-m^2)}{256\pi^3 m^2 M^2}
  (\Delta-2\ln\frac{M^2-m^2}{m\,M}+2) \nonumber \\
  +\hspace{-3mm}&&\frac{\theta[M-2m]\,\theta[3m-M]}{256\pi^3 m^2 M^2}\Bigl{[}
  (M^2-m^2-M\sqrt{M^2-4m^2})\Delta+2(M^2-m^2)(1-\ln\frac{M^2-m^2}{m\,M}) \nonumber \\
  +\hspace{-3mm}&&M\sqrt{M^2-4m^2}(\ln\frac{M^2-4m^2}{m^2}+\frac{\pi}{\sqrt{3}}-4) \Bigr{]}
  \nonumber \\
  +\hspace{-3mm}&&\frac{\theta[M-3m]}{256\pi^3 m^2 M^2}\Bigl{[} (M^2-m^2-M\sqrt{M^2-4m^2})
  \Delta+M\sqrt{M^2-4m^2}(\ln\frac{M^2-4m^2}{m^2}+\frac{\pi}{\sqrt{3}}-4)
  \nonumber \\
  +\hspace{-3mm}&&2(M^2-m^2)(1-\ln\frac{M^2-m^2}{m\,M})
  +2M^2\int_{y_1}^{y_2}\left( F-\sqrt{3}\,tg^{-1}\frac{F}{\sqrt{3}}\right)d\,y \Bigr{]}\,,
\eeqa
where
\beqa
  \Delta&=&\frac{2}{\epsilon}+2\ln4\pi-2\gamma-2\ln\frac{m^2}{\mu^2}\,, \nonumber \\
  F&=&\biggl{(}\frac{M^2 y^2-M^2 y-3m^2 y+4m^2}{y(M^2 y-M^2+m^2)}\biggr{)}^{1/2}\,, \nonumber \\
  y_{1,2}&=&\frac{M^2+3m^2\mp(M^4-10m^2 M^2+9m^4)^{1/2}}{2M^2}\,.
\eeqa
On the other hand, using the ameliorated cutting rules we also have
\beqa
  &&Im\,\mu^{2\epsilon}\int\frac{d^D k_1}{(2\pi)^D}\frac{d^D k_2}{(2\pi)^D}
  \frac{1}{k_1^2+i\varepsilon}\frac{1}{k_2^2-m^2+i\varepsilon}
  \frac{1}{(k_2-k_1)^2-m^2+i\varepsilon}\frac{1}{k_1^2-m^2+i\varepsilon}
  \frac{1}{(k_1-p)^2-m^2+i\varepsilon} \nonumber \\
  =\hspace{-3mm}&&Im\,\mu^{2\epsilon}P\int\frac{d^D k_1}{(2\pi)^D}\frac{d^D k_2}{(2\pi)^D}
  \Bigl{[} \frac{4i\,\pi^3\,\theta[k_{10}]\delta(k_1^2-m^2)\theta[k_{20}]\delta(k_2^2-m^2)
  \theta[p_0-k_{10}-k_{20}]\delta((p-k_1-k_2)^2-m^2)}{(p-k_1)^2((p-k_1)^2-m^2)} \nonumber \\
  -\hspace{-3mm}&&\frac{2\pi^2\theta[k_{10}]\delta(k_1^2)\theta[p_0-k_{10}]\delta((p-k_1)^2-m^2)}
  {(k_2^2-m^2)((k_2-k_1)^2-m^2)(k_1^2-m^2)}-
  \frac{2\pi^2\theta[k_{10}]\delta(k_1^2-m^2)\theta[p_0-k_{10}]\delta((p-k_1)^2-m^2)}
  {k_1^2(k_2^2-m^2)((k_2-k_1)^2-m^2)} \Bigr{]} \nonumber \\
  =\hspace{-3mm}&&\frac{1}{64\pi^3}P\int_0^{\infty}k_1^2\,d\,k_1
  \int_0^{\infty}k_2^2\,d\,k_2\int_{-1}^1 d\,x\frac{\delta(M-\sqrt{m^2+k_1^2}-
  \sqrt{m^2+k_2^2}-\sqrt{m^2+k_1^2+k_2^2+2 k_1 k_2\,x})}{\sqrt{m^2+k_1^2}
  \sqrt{m^2+k_2^2}\sqrt{m^2+k_1^2+k_2^2+2 k_1 k_2\,x}} \nonumber \\
  \times\hspace{-3mm}&&\frac{1}{(M^2+m^2-2M\sqrt{m^2+k_1^2})(M^2-2M\sqrt{m^2+k_1^2})} \nonumber \\
  +\hspace{-3mm}&&\frac{2\pi^2\mu^{2\epsilon}\Gamma(\epsilon/2)}{(4\pi)^{D/2}m^2}\int
  \frac{d^D k_1}{(2\pi)^D}\theta[k_{10}]\delta(k_1^2)\theta[p_0-k_{10}]\delta((p-k_1)^2-m^2) \nonumber \\
  -\hspace{-3mm}&&\frac{2\pi^2\mu^{2\epsilon}\Gamma(\epsilon/2)}{(4\pi)^{D/2}m^2}\int_0^1
  \frac{d\,x}{(x^2-x+1)^{\epsilon/2}}\int\frac{d^D k_1}{(2\pi)^D}
  \theta[k_{10}]\delta(k_1^2-m^2)\theta[p_0-k_{10}]\delta((p-k_1)^2-m^2) \nonumber \\
  =\hspace{-3mm}&&\frac{\theta[M-3m]m^2}{64\pi^3 M}\int_1^{\frac{M^3-3m^2}{2m\,M}}
  \frac{\sqrt{(x^2-1)(M^2-2m\,M\,x-3m^2)}}{(M^2+m^2-2m\,M\,x)^{3/2}(M-2m\,x)}
  d\,x \nonumber \\
  +\hspace{-3mm}&&\frac{\theta[M-m](M^2-m^2)}{256\pi^3 m^2 M^2}(\Delta-2\ln\frac{M^2-m^2}{m\,M}+2)
  \nonumber \\
  -\hspace{-3mm}&&\frac{\theta[M-2m]\sqrt{M^2-4m^2}}{256\pi^3 m^2 M}
  (\Delta-\ln\frac{M^2-4m^2}{m^2}+4-\frac{\pi}{\sqrt{3}})\,.
\eeqa
Through numerical calculations we find that Eq.(13) is equal to Eq.(11), as shown in Fig.7.
\begin{figure}[htbp]
\begin{center}
  \epsfig{file=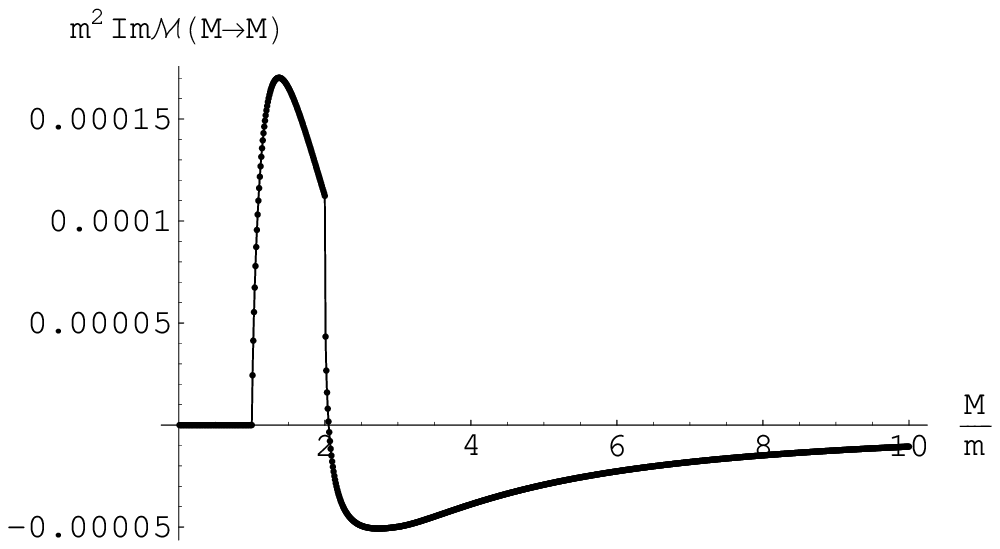,width=7.5cm} \\
  \caption{Results of Eq.(13) (plotted as dot) and Eq.(11) (plotted as line).}
\end{center}
\end{figure}
We note that a similar discussion has been present in Ref.\cite{c3}.

All of the four examples show that the ameliorated cutting rules agrees with the conventional integral algorithm very well. There are also many similar examples which we don't list here.

\section{Breaking of the optical theorem}

It's well-known that the optical theorem has put a strong constraint on the imaginary part of physical amplitude. Does it really right? In this section we will discuss this problem.

We firstly simply review the origin of the optical theorem. It's well known that the S-matrix is unitary: $S^{\dagger}S=1$. Inserting $S=1+i T$ one has:
\beq
  -i(T-T^{\dagger})\,=\,T^{\dagger}T\,.
\eeq
Take the matrix element of this equation between the two-particle states $|{\bf p}_1{\bf p}_2\hspace{-1mm}>$ and $|{\bf k}_1{\bf k}_2\hspace{-1mm}>$. To evaluate the right-hand side, insert a complete set of intermediate states,
\beq
  <\hspace{-1mm}{\bf p}_1{\bf p}_2|T^{\dagger}T|{\bf k}_1{\bf k}_2\hspace{-1mm}>\,=\,
  \sum_f\int d\prod_f <\hspace{-1mm}{\bf p}_1{\bf p}_2|T^{\dagger}|f\hspace{-1mm}>
  <\hspace{-1mm}f|T|{\bf k}_1{\bf k}_2\hspace{-1mm}>\,,
\eeq
where the sum runs over all possible sets $f$ of intermediate states and $d\prod_f$ represents the phase space integral of $f$. Now express the T-matrix elements as invariant matrix elements $\cal M$ times 4-momentum-conserving delta functions. Eq.(14) then becomes
\beq
  {\cal M}(k_1 k_2\rightarrow p_1 p_2)-{\cal M}^{\ast}(p_1 p_2\rightarrow k_1 k_2)\,=\,
  i\sum_f \int d\prod_f{\cal M}^{\ast}(p_1 p_2\rightarrow f)
  {\cal M}(k_1 k_2\rightarrow f)\,.
\eeq
Setting $k_i=p_i$ and applying the kinematic factors required by Eq.(16) to build a cross section, one obtains the standard form of the optical theorem,
\beq
  Im{\cal M}(k_1 k_2\rightarrow k_1 k_2)\,=\,2 E_{cm} p_{cm}
  \sigma_{tot}(k_1 k_2\rightarrow anything)\,,
\eeq
where $E_{cm}$ is the total center-of-mass energy and $p_{cm}$ is the momentum of either particle in the center-of-mass frame. Considering the $1\rightarrow 1$ process one obtains the well-known relationship,
\beq
  Im{\cal M}(a\rightarrow a)\,=\,m_a\Gamma_a\,,
\eeq
where $m_a$ and $\Gamma_a$ is the mass and decay width of particle $a$.

It seems that the optical theorem is a straightforward consequence of the unitarity of the S-matrix. But we find that there are some severe contradictions in the optical theorem. We firstly discuss a concrete physical amplitude. It is known that the imaginary part of physical amplitude comes from two cases: one is the coupling constant has imaginary part, the other is the Feynman amplitude has branch cut. At one-loop level a physical amplitude can be expressed as
\beq
  {\cal M}(i\rightarrow f)\,=\,\sum_k g_k (a_k+i\,b_k)\,,
\eeq where the sum runs over different interactions, $a_k$ is a
real number, $b_k$ is the imaginary part coming from the branch
cut of Feynman amplitude, and all of the imaginary parts of the
coupling constants are included in the coefficient $g_k$. For
convenience we define two concepts: the {\em quasi-imaginary part}
and the {\em quasi-real part} of Feynman amplitude, which
separately represent the parts of Feynman amplitude coming from
the branch cut and normal integral. For example in Eq.(19) the
{\em quasi-imaginary part} of the physical amplitude is $\sum_k
g_k b_k$ and the {\em quasi-real part} of the physical amplitude
is $\sum_k g_k a_k$. Under the charge conjugation transformation
only the imaginary part of the coupling constant changes into its
contra-value, all of the other quantities in Eq.(19) keep
unchanged. So we have from the CPT conservation law \beq
  {\cal M}^{\ast}(f\rightarrow i)\,=\,{\cal M}^{\ast}(\bar{i}\rightarrow \bar{f})
  \,=\,\sum_k g_k(a_k-i\,b_k)\,.
\eeq
From Eqs.(19) and (20) we have
\beq
  {\cal M}(i\rightarrow f)-{\cal M}^{\ast}(f\rightarrow i)\,=\,2\,i\,\sum_k g_k b_k\,\equiv\,
  2\,i\,\tilde{Im}{\cal M}(i\rightarrow f)\,,
\eeq
where $\tilde{Im}$ takes the {\em quasi-imaginary part} of the physical amplitude. Obviously this result clearly demonstrates the relationship between the optical theorem and the cutting rules. If the optical theorem is right, one will have for the physical amplitude of top quark decaying into charm quark and gauge boson Z
\beq
  \tilde{Im}{\cal M}(t\rightarrow c\,Z)\,=\,\half\sum_f \int d\prod_f {\cal M}^{\ast}(c\,Z\rightarrow f)
  {\cal M}(t\rightarrow f)\,.
\eeq
On the other hand, one can calculate the {\em quasi-imaginary part} of ${\cal M}(t\rightarrow c\,Z)$ from the branch cuts of the Feynman amplitude. The one-loop Feynman diagrams of $t\rightarrow c\,Z$ have been shown in Fig.8.
\begin{figure}[htbp]
\begin{center}
  \epsfig{file=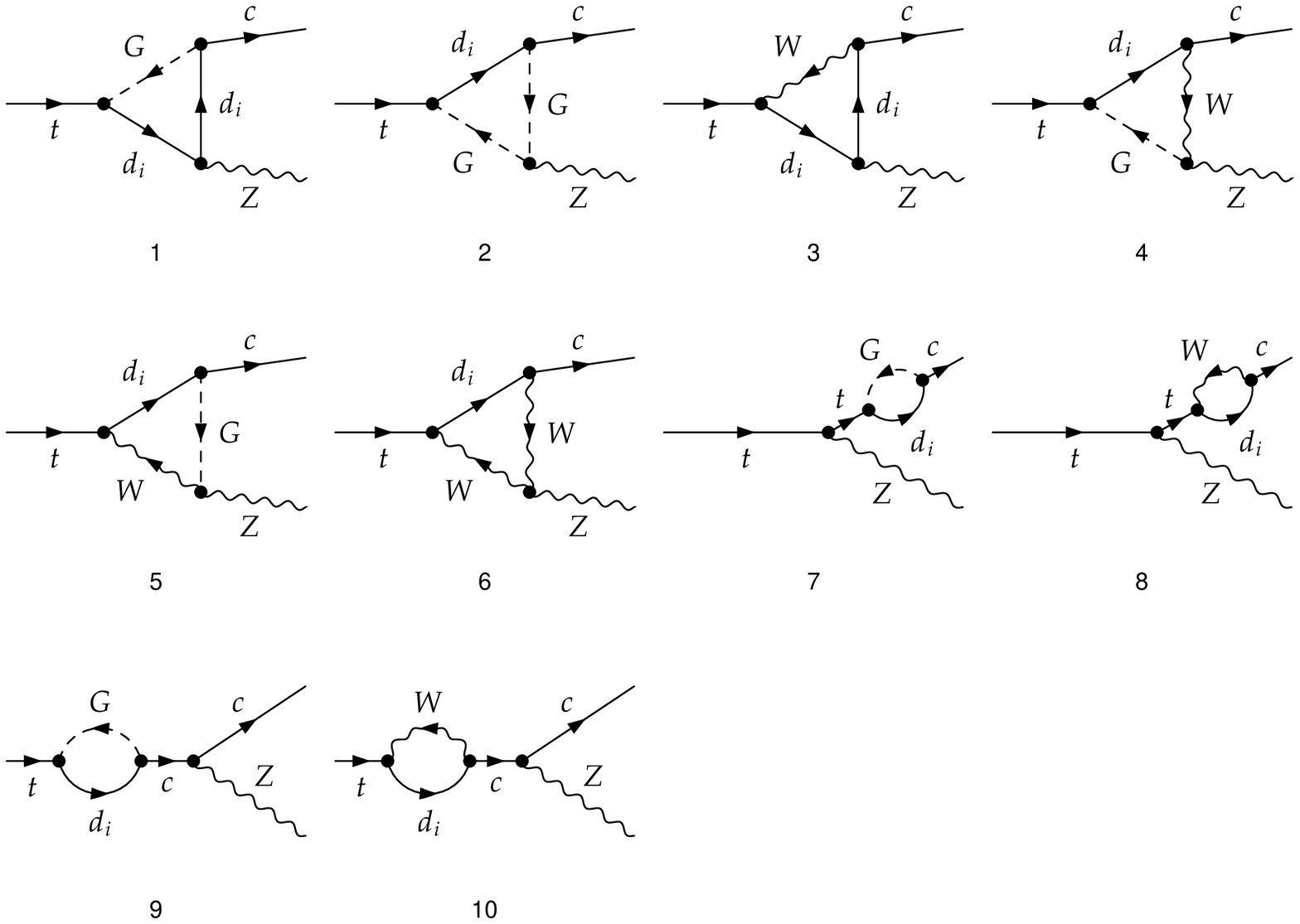,width=12cm} \\
  \caption{One-loop diagrams of the physical amplitude of $t\rightarrow c\,Z$.}
\end{center}
\end{figure}
We note that there is no need to introduce counterterm into the one-loop physical amplitude of $t\rightarrow c\,Z$ after including the contribution of the amplitude of Fig.8. From the branch cuts of Fig.8 we find a contradiction in the optical theorem: Eq.(22) forbids the cuts representing the contribution of $\sum_f\int d\prod_f {\cal M}^{\ast}(c\rightarrow f){\cal M}(t\rightarrow Z\,f)$ in which the cuts of $t\rightarrow c$ two-point diagrams in the seven and eight diagrams of Fig.8 are included, but that cuts of $t\rightarrow c$ two-point diagrams in the seven and eight diagrams of Fig.8 are permitted by Eq.(18). Obviously the contribution of the cuts of $t\rightarrow c$ two-point diagrams in the seven and eight diagrams of Fig.8 to the physical amplitude isn't equal to zero.

In fact there exists a more severe contradiction. From the
unitarity of S-matrix we also have $T^{\dagger}T=T T^{\dagger}$.
Thus Eq.(15) can be changed into \beq
  <\hspace{-1mm}{\bf p}_1{\bf p}_2|T^{\dagger}T|{\bf k}_1{\bf k}_2\hspace{-1mm}>\,=\,
  \sum_f\int d\prod_f <\hspace{-1mm}{\bf p}_1{\bf p}_2|T|f\hspace{-1mm}>
  <\hspace{-1mm}f|T^{\dagger}|{\bf k}_1{\bf k}_2\hspace{-1mm}>\,.
\eeq
Since each intermediate state $f$ has its own unique parameters (e.g. mass), Eq.(23) being equal to Eq.(15) means all of the terms about $f$ are separately equal, i.e. for arbitrary $f$
\beq
  <\hspace{-1mm}{\bf p}_1{\bf p}_2|T^{\dagger}|f\hspace{-1mm}>\,=\,
  <\hspace{-1mm}{\bf p}_1{\bf p}_2|T|f\hspace{-1mm}>\,, \hspace{10mm}
  <\hspace{-1mm}f|T|{\bf k}_1{\bf k}_2\hspace{-1mm}>\,=\,
  <\hspace{-1mm}f|T^{\dagger}|{\bf k}_1{\bf k}_2\hspace{-1mm}>\,.
\eeq
Since $f$ are arbitrary, we ultimately obtain
\beq
  {\cal M}(k_1 k_2\rightarrow p_1 p_2)-{\cal M}^{\ast}(p_1 p_2\rightarrow k_1 k_2)\,=\,
  0\,.
\eeq
Obviously Eq.(25) contradicts Eq.(16).

Another problem about the optical theorem is: if the charge
conjugation is conserved, from the CPT conservation law the
left-hand side of Eq.(16) is a pure imaginary, but the right-hand
side of Eq.(16) may contain real part which comes from the branch
cut of the physical amplitudes. Why are there such contradictions
and problem? In fact the S-matrix is only a symbolistic operator,
it doesn't have the algebraic algorithm that a true operator
should have, e.g. the multiplication and addition. This is because
the S-matrix represents the summation of the interactions of the
physical world, so it is no possible to express it by a concrete,
having certain algebraic algorithm's, operator. Therefore the
derivation of Eqs.(14-16) isn't right.

\section{Gauge independence of physical result under ameliorated cutting rules}

In order to investigate whether the ameliorated cutting rules is reasonable, we calculate the physical amplitude $t\rightarrow c\,Z$ to see if the ameliorated cutting rules keeps the physical result gauge independent.

In Fig.8 we have shown the one-loop diagrams of the physical process $t\rightarrow c\,Z$. Since there is no tree-level diagram, it is no need to introduce counterterm to one-loop level. We note that in the following calculations we have used the program packages {\em FeynArts, FeynCalc, FormCalc and LoopTools} \cite{c6}. Our calculations have shown that the {\em quasi-real part} of ${\cal M}(t\rightarrow c\,Z)$ is gauge-parameter independent. So we only need to consider the {\em quasi-imaginary part} of ${\cal M}(t\rightarrow c\,Z)$. Using the ameliorated cutting rules we obtain (see also Fig.1)
\beq
  \tilde{Im}{\cal M}(t\rightarrow c\,Z)|_{\xi}\,=\,0\,,
\eeq
where subscript $\xi$ takes the gauge-dependent part of the quantity. From Eq.(26) one can easily see that all of the unphysical cuts (e.g. $t\rightarrow G^{+}d_i$) in Fig.8 have been cancelled out in the physical amplitude by the ameliorated cutting rules. This coincides with the conventional knowledge \cite{c7}. Obviously such result makes the decay width of $t\rightarrow c\,Z$ gauge independent to two-loop level.

By the way, from the above discussions we know that the optical theorem gives a contradictory result about the {\em quasi-imaginary part} of ${\cal M}(t\rightarrow c\,Z)$. If keeping the cuts of $t\rightarrow c$ two-point diagrams in the seven and eight diagrams of Fig.8 and for the first six diagrams of Fig.8 only keeping the first kind cut of Fig.1 required by Eq.(22), we obtain
\beqa
  \tilde{Im}{\cal M}(t\rightarrow c\,Z)|_{\xi}&=&\bar{c}\,
  {\xslash \epsilon^{\ast}}\gamma_L\,t\,\sum_i\frac{V_{2i}V^{\ast}_{3i}\,e^3
  (3-4 s_W^2)(x_c-\xi_W-x_{d,i})}{384\pi\,c_W\,s_W^3 x_c} \nonumber \\
  &\times&\left( x_c^2-2(\xi_W+x_{d,i})x_c+(\xi_W-x_{d,i})^2 \right)^{1/2}\,
  \theta[m_c-m_{d,i}-M_W\sqrt{\xi_W}]\,,
\eeqa
where $\gamma_L$ is the left-handed helicity operator, $V_{2i}$ and $V_{3i}$ are the CKM matrix elements \cite{c8}, $e$ is electron charge, $s_W$ and $c_W$ is the sine and cosine of the weak mixing angle, $M_W$ and $\xi_W$ is the mass and gauge parameter of gauge boson $W$, $m_c$ and $m_{d,i}$ are the masses of charm quark and down-type $i$ quark, $x_c=m_c^2/M_W^2$ and $x_{d,i}=m_{d,i}^2/M_W^2$. This result is gauge dependent and will lead to the decay width of $t\rightarrow c\,Z$ gauge dependent. One can easily see that the gauge-dependent terms of Eq.(27) come from the unphysical cuts of $t\rightarrow c$ two-point diagrams in the seven and eight diagrams of Fig.8.

\section{Conclusion}

In order to calculate the contribution of the singularity of Feynman propagator to Feynman amplitude we investigate the cutting rules and the optical theorem. We ameliorate the cutting rules in order to make it suitable for actual calculation and give the right result of the imaginary part of physical amplitude. The calculations of the imaginary parts of several Feynman diagrams show that the ameliorated cutting rules agrees with the conventional integral algorithm very well (see Fig.6,7 and Eq.(6,8,9,10)). On the other hand, through careful investigation we find that the optical theorem has severe contradictions and problem thus isn't right. Besides, the calculation of the physical amplitude $t\rightarrow c\,Z$ shows that the ameliorated cutting rules keeps the decay width of the physical process gauge independent to two-loop level (see Eq.(26)).

In the viewpoint of the conventional cutting rules the branch cut of Feynman diagram contributes the imaginary part to physical amplitude. Our investigation further finds that although the Feynman diagram doesn't represent the real physical process, it does provide a physical-like picture to describe the on-shell virtual particles' propagating processes happening in the quantum field vacuum if such processes exist. The ameliorated cutting rules not only gives a feasible algorithm to calculate the contributions of the virtual processes to physical amplitude, but also can help us to discover the deep-seated physical meaning of the quantum field theory.

\vspace{5mm} {\bf \Large Acknowledgments} \vspace{2mm}

The author thanks Prof. Cai-dian Lu for the fruitful discussions and the corrections of the words. The author also thanks Yue-long Shen, doctor Jian-feng Cheng, Xian-qiao Yu, Ge-liang Song and Ying Li for the fruitful discussions.

\end{document}